\documentclass [a4paper,12pt] {article}

\usepackage{latexsym}
\usepackage{amsfonts}
\usepackage{amsmath}
\usepackage{amssymb}

  \def\pp{{\mathchoice
          {%displaystyle
              \kern 1pt%
              \raise 1pt
              \vbox{\hrule width5pt height0.4pt depth0pt
                    \kern -2pt
                    \hbox{\kern 2.3pt
                          \vrule width0.4pt height6pt depth0pt
                          }
                    \kern -2pt
                    \hrule width5pt height0.4pt depth0pt}%
                    \kern 1pt
           }
            {%textstyle
              \kern 1pt%
              \raise 1pt
              \vbox{\hrule width4.3pt height0.4pt depth0pt
                    \kern -1.8pt
                    \hbox{\kern 1.95pt
                          \vrule width0.4pt height5.4pt depth0pt
                          }
                    \kern -1.8pt
                    \hrule width4.3pt height0.4pt depth0pt}%
                    \kern 1pt
            }
            {%scriptstyle
              \kern 0.5pt%
              \raise 1pt
              \vbox{\hrule width4.0pt height0.3pt depth0pt
                    \kern -1.9pt  %[e]=0.15pt
                    \hbox{\kern 1.85pt
                          \vrule width0.3pt height5.7pt depth0pt
                          }
                    \kern -1.9pt
                    \hrule width4.0pt height0.3pt depth0pt}%
                    \kern 0.5pt
            }
            {%scriptscriptstyle
              \kern 0.5pt%
              \raise 1pt
              \vbox{\hrule width3.6pt height0.3pt depth0pt
                    \kern -1.5pt
                    \hbox{\kern 1.65pt
                          \vrule width0.3pt height4.5pt depth0pt
                          }
                    \kern -1.5pt
                    \hrule width3.6pt height0.3pt depth0pt}%
                    \kern 0.5pt%}
            }
        }}

  \def\mm{{\mathchoice
                       {%displaystyle
                             \kern 1pt
               \raise 1pt    \vbox{\hrule width5pt height0.4pt depth0pt
                                  \kern 2pt
                                  \hrule width5pt height0.4pt depth0pt}
                             \kern 1pt}
                       {%textstyle
                            \kern 1pt
               \raise 1pt \vbox{\hrule width4.3pt height0.4pt depth0pt
                                  \kern 1.8pt
                                  \hrule width4.3pt height0.4pt depth0pt}
                             \kern 1pt}
                       {%scriptstyle
                            \kern 0.5pt
               \raise 1pt
                            \vbox{\hrule width4.0pt height0.3pt depth0pt
                                  \kern 1.9pt
                                  \hrule width4.0pt height0.3pt depth0pt}
                            \kern 1pt}
                       {%scriptscriptstyle
                           \kern 0.5pt
             \raise 1pt  \vbox{\hrule width3.6pt height0.3pt depth0pt
                                  \kern 1.5pt
                                  \hrule width3.6pt height0.3pt depth0pt}
                           \kern 0.5pt}
                       }}

\catcode`@=11
\def\un#1{\relax\ifmmode\@@underline#1\else
        $\@@underline{\hbox{#1}}$\relax\fi}
\catcode`@=12

\let\du=\du                     % dot-under
\def\a{\alpha}
\def\b{\beta}
\def\c{\chi}
\def\d{\delta}

\def\f{\phi}
\def\g{\gamma}

\def\m{\mu}
\def\n{\nu}

\def\q{\theta}

\def\s{\sigma}

\def\x{\xi}

\def\F{\Phi}

\def\ve{\varepsilon}

\def\bo{{\raise-.5ex\hbox{\large$\Box$}}}               % D'Alembertian
\def\pa{\partial}                                       % curly d
\def\TH{{\raise.2ex\hbox{$\displaystyle \bigodot$}\mskip-4.7mu \llap H \;}}
\def\face{{\raise.2ex\hbox{$\displaystyle \bigodot$}\mskip-2.2mu \llap {$\ddot
        \smile$}}}                                      % happy face
\def\Tilde#1{\widetilde{#1}}                    % big tilde
\def\Bar#1{\overline{#1}}                       % big bar
\def\leftrightarrowfill{$\mathsurround=0pt \mathord\leftarrow \mkern-6mu
        \cleaders\hbox{$\mkern-2mu \mathord- \mkern-2mu$}\hfill
        \mkern-6mu \mathord\rightarrow$}
\def\dvec#1{\vbox{\ialign{##\crcr
        \leftrightarrowfill\crcr\noalign{\kern-1pt\nointerlineskip}
        $\hfil\displaystyle{#1}\hfil$\crcr}}}           % <--> accent
\def\dt#1{{\buildrel {\hbox{\LARGE .}} \over {#1}}}     % dot-over for sp/sb
\def\frac#1#2{{\textstyle{#1\over\vphantom2\smash{\raise.20ex
        \hbox{$\scriptstyle{#2}$}}}}}                   % fraction
\def\sfrac#1#2{{\vphantom1\smash{\lower.5ex\hbox{\small$#1$}}\over
        \vphantom1\smash{\raise.4ex\hbox{\small$#2$}}}} % alternate fraction
\def\bfrac#1#2{{\vphantom1\smash{\lower.5ex\hbox{$#1$}}\over
        \vphantom1\smash{\raise.3ex\hbox{$#2$}}}}       % "
\def\afrac#1#2{{\vphantom1\smash{\lower.5ex\hbox{$#1$}}\over#2}}    % "
\def\on#1#2{\mathop{\null#2}\limits^{#1}}               % arbitrary accent
\def\bvec#1{\on\leftarrow{#1}}                  % backward vector accent
\def\[{\lfloor{\hskip 0.35pt}\!\!\!\lceil}
\def\]{\rfloor{\hskip 0.35pt}\!\!\!\rceil}

\def\du#1#2{_{#1}{}^{#2}}

\def\fracm#1#2{\hbox{\large{${\frac{{#1}}{{#2}}}$}}}
\def\ha{{\fracmm12}}

\def\un{\underline}
\def\fracmm#1#2{{{#1}\over{#2}}}

\def\low#1{{\raise -3pt\hbox{${\hskip 0.75pt}\!_{#1}$}}}

\def\Dot#1{\buildrel{_{_{\hskip 0.01in}\bullet}}\over{#1}}
\def\dt#1{\Dot{#1}}

\def\Tilde#1{{\widetilde{#1}}\hskip 0.015in}

\newskip\humongous \humongous=0pt plus 1000pt minus 1000pt
\def\caja{\mathsurround=0pt}
\def\eqalign#1{\,\vcenter{\openup2\jot \caja
        \ialign{\strut \hfil$\displaystyle{##}$&$
        \displaystyle{{}##}$\hfil\crcr#1\crcr}}\,}
\newif\ifdtup

\parskip=\medskipamount                 % space between paragraphs (LaTeX)
\thicklines                         % thick straight lines for pictures (LaTeX)

\begin{document}
\thispagestyle{empty}

{\hbox to\hsize{
\vbox{\noindent May 2005 \hfill hep-th/0504191
{} \hfill \\revised version }}}

\noindent
\vskip1.3cm
\begin{center}

{\Large\bf Summing up Non-anti-commutative K\"ahler \vglue.1in
 Potential~\footnote{
Supported in part by the Japanese Society for Promotion of Science (JSPS)}}
\vglue.2in

Tomoya Hatanaka~\footnote{Email address: thata@kiso.phys.metro-u.ac.jp},
Sergei V. Ketov~\footnote{Email address: ketov@phys.metro-u.ac.jp},
and Shin Sasaki~\footnote{Email address: shin-s@phys.metro-u.ac.jp}

{\it Department of Physics\\
     Tokyo Metropolitan University\\
     1--1 Minami-osawa, Hachioji-shi\\
     Tokyo 192--0397, Japan}
\end{center}
\vglue.2in
\begin{center}
{\Large\bf Abstract}
\end{center}

We offer a simple non-perturbative formula for the component action of a 
generic N=1/2 supersymmetric chiral model in terms of an arbitrary 
number of chiral superfields in four dimensions, which is obtained by the 
Non-Anti-Commutative (NAC) deformation of a generic four-dimensional N=1
supersymmetric Non-Linear Sigma-Model (NLSM) described by arbitrary K\"ahler
superpotential and scalar superpotential. The auxiliary integrations 
responsible for fuzziness are eliminated in the case of a single chiral 
superfield. The scalar potential in components is derived by eliminating the 
auxiliary fields. The NAC-deformation of the $CP^1$ K\"ahler NLSM with an 
arbitrary scalar superpotential is calculated as an example.
  
\newpage

\section{Introduction}

There was a lot of recent activity in investigating various aspects of 
Non-Anti-Commutative (NAC) superspace and related deformations of 
supersymmetric field theories (see, e.g., the most recent references 
\cite{sei,ital,chandra,buch,nac1,spain,nbi} directly related to our title, and
the references therein for the earlier work in the NAC-deformed N=1 
superspace). It is supposed to enhance our understanding of the role of 
spacetime in supersymmetry, while keeping globally supersymmetric field theory
 under control. 

We work in four-dimensional Euclidean~\footnote{The use of 
Atiyah-Ward spacetime of the signature $(+,+,-,-)$ is another possibility
\cite{kgn}.} N=1 superspace
$(x^{\m},\q^{\a},\bar{\q}^{\dt{\a}})$, and use the standard notation \cite{wb}.
The NAC deformation is given by  
$$ \{ \q^{\a},\q^{\b} \}_*=C^{\a\b}~~,
\eqno(1.1)$$
where $C^{\a\b}$ are some constants. The remaining superspace coordinates in 
the chiral basis ($y^{\m}=x^{\m}+i\q\s^{\m}\bar{\q}$, $\m,\n=1,2,3,4$ and 
$\a,\b,\ldots=1,2$) still (anti)commute,
$$\[y^{\m},y^{\n}\]= \{ \bar{\q}^{\dt{\a}},\bar{\q}^{\dt{\b}} \}=
\{ \q^{\a},\bar{\q}^{\dt{\b}} \}=\[y^{\m},\q^{\a}\]=
\[y^{\m},\bar{\q}^{\dt{\a}}\]=0~.\eqno(1.2)$$

The $C^{\a\b}\neq 0$ explicitly break the four-dimensional `Lorentz' 
invariance at the fundamental level. The NAC nature of $\q$'s can be fully 
taken into account by using the Moyal-Weyl-type (star) product of
superfields \cite{sei}~,
 $$ f(\q)* g(\q)=f(\q)\,
\exp\left(-\fracmm{C^{\a\b}}{2}\fracmm{\bvec{\pa}}{\pa
\q^{\a}}\fracmm{\vec{\pa}}{\pa\q^{\b}}\right)g(\q)~, \eqno(1.3)$$
which respects the N=1 superspace chirality. The star product (1.3) is 
polynomial in the deformation parameter~,
$$ f(\q)*g(\q)=fg +(-1)^{{\rm deg}f}\fracmm{C^{\a\b}}{2}
\fracmm{\pa f}{\pa\q^{\a}}\fracmm{\pa g}{\pa\q^{\b}}-\det\,C
\fracmm{\pa^2 f}{\pa\q^2}\fracmm{\pa^2 g}{\pa\q^2}~~,\eqno(1.4)$$
where we have used the identity
$$ \det C = \fracm{1}{2}\ve_{\a\g}\ve_{\b\d}C^{\a\b}C^{\g\d}~,\eqno(1.5)$$
and the notation
$$ \fracmm{\pa^2}{\pa\q^2}= \fracm{1}{4}\ve^{\a\b}\fracmm{\pa}{\pa\q^{\a}}
\fracmm{\pa}{\pa\q^{\b}}~~.\eqno(1.6)$$

We also use the following book-keeping notation for 2-component spinors:
$$ \q\c=\q^{\a}\c_{\a}~,\quad  \bar{\q}\bar{\c}=\bar{\q}_{\dt{\a}}
\bar{\c}^{\dt{\a}}~,\quad \q^2= \q^{\a}\q_{\a}~,\quad 
\bar{\q}^2= \bar{\q}_{\dt{\a}}\bar{\q}^{\dt{\a}}.\eqno(1.7)$$
The spinorial indices are raised and lowered by the use of two-dimensional 
Levi-Civita symbols \cite{wb}. Grassmann integration amounts to Grassmann 
differentiation. The anti-chiral covariant derivative in the chiral superspace
basis is $\bar{D}_{\dt{\a}}=-\pa/ \pa \bar{\q}^{\dt{\a}}$. The field component 
expansion of a chiral superfield $\F$ reads 
$$ \F(y,\q)= \f(y) +\sqrt{2}\q\c(y)+\q^2 M(y)~~.\eqno(1.8)$$
An anti-chiral superfield $\Bar{\F}$ in the chiral basis is given by
$$\eqalign{
\Bar{\F}(y^{\m}-2i\q\s^{\m}\bar{\q},\bar{\q})=  ~ & ~ 
\bar{\f}(y) + \sqrt{2}\bar{\q}\bar{\c}(y) +\bar{\q}^2\bar{M}(y)  \cr
 ~ & ~ +\sqrt{2}\q\left( i\s^{\m}\pa_{\m}\bar{\c}(y)\bar{\q}^2-i\sqrt{2}\s^{\m}
\bar{\q}\pa_{\m}\bar{\f}(y)\right)+\q^2\bar{\q}^2\bo\bar{\f}(y)~,\cr}
\eqno(1.9)$$
where $\bo =\pa_{\m}\pa^{\m}$. The bars over fields serve to distinguish 
between the `left' and `right' components that are truly independent in 
Euclidean spacetime.
 
Our major concern in this Letter is a derivation of the NAC deformation of a
generic four-dimensional N=1 supersymmetric action 
$$ S = \int d^4 x \left[ \int d^2\q d^2\bar{\q}\, K(\F^i,\Bar{\F}{}^{\bar{j}})+
\int d^2\q\, W(\F^i) + \int  d^2\bar{\q}\,\Bar{W}(\Bar{\F}{}^{\bar{j}})\right]
\eqno(1.10)$$
specified by a K\"ahler superpotential $K(\F,\Bar{\F})$ and a scalar 
superpotential $W(\F)$, in terms of an arbitrary number $n$  of chiral and
anti-chiral superfields, $i,\bar{j}=1,2,\ldots,n$. This problem in four 
dimensions was addressed in refs.~\cite{buch,nbi}, where the perturbative 
answers (in terms of infinite sums) were found. A similar problem in two 
dimensions was solved perturbatively in ref.~\cite{chandra}, while the 
non-perturbative summation (in terms of finite functions) was done in 
ref.~\cite{spain}. In this Letter we give simple non-perturbative formulas in 
four dimensions and offer a simple way of their derivation. Our results 
presumably amount to a full summation of the inifinite sums in 
refs.~\cite{buch,nbi}. Summing up is crucial for further non-perturbative 
physical applications of the NAC-deformation and its geometrical 
interpretation. We also made progress in eliminating the auxiliary 
integrations  and solving the auxiliary field 
equations, as well as in investigating some concrete examples (see below).
  
We use the chiral basis, which is most suitable for investigating  
NAC-deformation, and reduce the most non-trivial problem of calculation of the
 NAC-deformed K\"ahler superpotential to that for the NAC-deformed scalar 
superpotential. The remarkably simple non-trivial results about the 
NAC-deformed scalar superpotential are already available in 
refs.~\cite{nac1,spain}. In sect.~2 we describe our idea in the undeformed
case (it is not really new there). In sect.~3 we present the results of our 
calculation for the most general NAC-deformed action (1.10). In sect.~4 we 
specialize our results to the case of a single chiral superfield and the 
$CP^1$ (K\"ahler) superpotential, as the simplest non-trivial examples. 
Sect.~5 is our conclusion.

\section{Chiral reduction of K\"ahler superpotential}

Let's use the identity
$$\int d^4x d^2\q d^2\bar{\q}\,K(\F,\Bar{\F})=-\fracm{1}{4}
\int d^4y d^2\q \left.\bar{D}^2K(\F,\Bar{\F})\right|
\equiv \int d^4y \,L~~,\eqno(2.1)$$
where $|$ denotes the $\bar{\q}$-independent part of a superfield, and the
constraint $\bar{D}_{\dt{\a}}\F^i=0$. Having performed Grassmann 
differentiation in $\bar{D}^2K$, we arrive at the spacetime (NAC-undeformed) 
component Lagrangian in the chiral form, 
$$ L=\int d^2\q\,V\low{(K)}(\F^I)~,\eqno(2.2)$$
whose effective scalar superpotential $V\low{(K)}(\F^I)$ is given by 
$$V\low{(K)}= 
-\fracm{1}{2}K_{,\bar{p}\bar{q}}(\F(y,\q),\bar{\f}(y))\F^{\bar{p}\bar{q}}_{n+1}
 + K_{,\bar{p}}(\F(y,\q),\bar{\f}(y))\F^{\bar{p}}_{n+2}\eqno(2.3)$$
in terms of the {\it extended} set of the chiral superfields 
$\F^I=\{\F^i,\F^{\bar{p}\bar{q}}_{n+1},\F^{\bar{p}}_{n+2}\}$. 
We use the notation (valid for any function $F(\f,\bar{\f})$)
$$ F_{,i_1i_2\cdots i_s \bar{p}_1\bar{p}_2\cdots \bar{p}_t}=
\fracmm{\pa^{s+t}F}{\pa\f^{i_1}\pa\f^{i_2}\cdots\f^{i_s}
\pa\bar{\f}^{\bar{p}_1}\pa\bar{\f}^{\bar{p}_2}\cdots\pa
\bar{\f}^{\bar{p}_t}}~~~~,\eqno(2.4)$$
and the fact that $\left.\bar{\F}^{\bar{j}}\right|=\bar{\f}^{\bar{j}}(y)$.
 The additional (composite) $n(n+1)/2$ chiral superfields 
$\F^{\bar{p}\bar{q}}_{n+1}$ and $n$ chiral superfields $\F^{\bar{p}}_{n+2}$ 
are given by
$$ \F^{\bar{p}\bar{q}}_{n+1}(y,\q)=\fracm{1}{2}\left.(\bar{D}_{\dt{\a}}
\bar{\F}{}^{\bar{p}})(\bar{D}^{\dt{\a}}\bar{\F}{}^{\bar{q}})\right|=
\bar{\c}^{\bar{p}}\bar{\c}^{\bar{q}} -2\sqrt{2}i\left(\q\s^{\m}
\bar{\c}^{(\bar{p}}\right)\pa_{\m}\bar{\f}^{\bar{q})}
-2\q^2\pa^{\m}\bar{\f}^{(\bar{p}}\pa_{\m}\bar{\f}^{\bar{q})}
\eqno(2.5a)$$
and 
$$\F^{\bar{p}}_{n+2}(y,\q) =-\fracm{1}{4}\left.\bar{D}^2\bar{\F}{}^{\bar{p}}
\right|=\bar{M}^{\bar{p}}+\sqrt{2}i(\q\s^{\m}\pa_{\m}\bar{\c}^{\bar{p}})+
\q^2\bo\bar{\f}{}^{\bar{p}}~.\eqno(2.5b)$$ 
Therefore, the NAC deformation of the K\"ahler superpotential is not an
independent problem, since it is derivable from the NAC deformation of the
effective scalar superpotential (2.3) having the extended number of chiral 
superfields. 

The simple non-perturbative form of an arbitrary NAC-deformed scalar 
superpotential $V$, depending upon a single chiral superfield $\F$ of 
eq.~(1.8), was first calculated in ref.~\cite{nac1},
$$ \int d^2\q\,V_*(\F)=\fracm{1}{2c}\left[ V(\f+cM)-V(\f-cM)\right]
-\fracmm{\c^2}{4cM}\left[ V_{,\f}(\f+cM)- V_{,\f}(\f-cM)\right]~,\eqno(2.6)$$
where we have introduced the (finite) deformation parameter
$$ c=\sqrt{-\det C}~.\eqno(2.7)$$

As is clear from eq.~(2.6), the NAC-deformation in the single superfield case 
gives rise to the scalar potential split controlled by the auxiliary field $M$.
When using an identity
$$ f(x+a)-f(x-a) =a\fracmm{\pa}{\pa x}
\int^{+1}_{-1} d\x\,f(x+\x a)~,\eqno(2.8)$$
valid for any function $f$, we can rewrite eq.~(2.6) to the equivalent form,
$$ \int d^2\q\,V_*(\F)=\fracm{1}{2}M\fracmm{\pa}{\pa\f}\int^{+1}_{-1} d\x\,
V(\f+\x cM) -\fracmm{1}{4}\c^2\fracmm{\pa^2}{\pa\f^2}\int^{+1}_{-1} d\x\,
V(\f+\x cM)~,\eqno(2.9a)$$    
which is very suitable for an immediate generalization to the case of several
chiral superfields ({\sl cf.} ref.~\cite{spain}),
$$ \int d^2\q\,V_*(\F^I)=\fracm{1}{2}M^I\fracmm{\pa}{\pa\f^I}\Tilde{V}(\f,M)
-\fracmm{1}{4}(\c^I\c^J)\fracmm{\pa^2}{\pa\f^I\pa\f^J}\Tilde{V}(\f,M)~,
\eqno(2.9b)$$   
where the auxiliary pre-potential $\Tilde{V}$ has been introduced \cite{spain},
$$\Tilde{V}(\f,M)=\int^{+1}_{-1} d\x\,V(\f^I+\x cM^I)~.\eqno(2.10)$$
Therefore, the NAC-deformation of a generic scalar superpotential $V$ results
in its smearing or fuzziness controlled by the auxiliary fields $M^I$
 \cite{spain}.

\section{The NAC-deformed K\"ahler superpotential}

Our general result in components is just given by eq.~(2.9b) after a 
subsitution of eq.~(2.3) in terms of the definitions of sect.~2. We find
({\it cf.} refs.~\cite{chandra,buch,spain,nbi})
$$ \eqalign{
L~ = ~& ~ \fracm{1}{2} M^iY_{,i}+
\fracm{1}{2}\pa^{\m}\bar{\f}{}^{\bar{p}}\pa_{\m}\bar{\f}{}^{\bar{q}}
Z_{,\bar{p}\bar{q}} + \fracm{1}{2}\bo\bar{\f}{}^{\bar{p}}
Z_{,\bar{p}} -\fracm{1}{4}(\c^i\c^j)Y_{,ij} \cr
 ~& ~ -\fracm{1}{2}i(\c^i\s^{\m}\bar{\c}{}^{\bar{p}})\pa_{\m}
\bar{\f}{}^{\bar{q}}Z_{,i\bar{p}\bar{q}}
-\fracm{1}{2}i(\c^i\s^{\m}\pa_{\m}\bar{\c}{}^{\bar{p}})Z_{,i\bar{p}}~,\cr}
\eqno(3.1)$$
where we have introduced the (component) smeared K\"ahler pre-potential
$$ Z(\f,\bar{\f},M)=\int^{+1}_{-1}d\x\, K^{\x} \quad {\rm with}\quad
K^{\x}\equiv K(\f^i+\x cM^i,\bar{\f}{}^{\bar{j}})~~,\eqno(3.2a)$$
as well as the extra (auxiliary) pre-potential 
$$ Y(\f,\bar{\f},M,\bar{M})=\bar{M}{}^{\bar{p}}Z_{,\bar{p}}
 -\fracm{1}{2}(\bar{\c}{}^{\bar{p}}\bar{\c}{}^{\bar{q}})
Z_{,\bar{p}\bar{q}} +c\int^{+1}_{-1}d\x \x\left[ 
\pa^{\m}\bar{\f}{}^{\bar{p}}\pa_{\m}\bar{\f}{}^{\bar{q}}
K^{\x}_{,\bar{p}\bar{q}} +\bo\bar{\f}{}^{\bar{p}}K^{\x}_{,\bar{p}}\right]~.
\eqno(3.2b)$$
We verified that eq.~(3.1) reduces to the standard (K\"ahler) N=1 
supersymmetric non-linear sigma-model in the limit $c\to 0$. Also, in the case
of a free (bilinear) K\"ahler potential $K=\d_{i\bar{j}}\F^i\bar{\F}^{\bar{j}}
$, there is no deformation at all. 

Given, in addition, an independent scalar superpotential $W(\F)$, as in 
eq.~(1.10), the following component terms are to be added to eq.~(3.1):
$$ L_{\rm potential}=\fracm{1}{2}M^i\Tilde{W}_{,i}-\fracm{1}{4}(\c^i\c^j)
\Tilde{W}_{,ij}+ \bar{M}^{\bar{p}}\bar{W}_{,\bar{p}}-  
\fracm{1}{2}(\bar{\c}^{\bar{p}}\bar{\c}^{\bar{q}})\bar{W}_{,\bar{p}\bar{q}}~,
\eqno(3.3)$$
where we have used the fact that the anti-chiral scalar superpotential terms 
are inert under the NAC-deformation, and we have introduced the smeared scalar 
pre-potential \cite{spain}
$$ \Tilde{W}(\f,M)= \int^{+1}_{-1}d\x\, W(\f^i+\x cM^i)~~.\eqno(3.4)$$

Though our general results of this section are quite explicit and 
non-perturbative, as regards their applications, it is desirable to perform 
all integrations over $\x$ (thus evaluating the smearing effects) and 
eliminate the auxiliary fields  $(M,\bar{M})$ by using their algebraic 
equations of motion. It is clear that the $\x$-integrals can be evaluated 
once the K\"ahler and scalar functions, $K$ and $W$, are specified. The 
anti-chiral auxiliary fields $\bar{M}^{\bar{p}}$ enter the action linearly (as
Lagrange multipliers), so that their algebraic equations of motion determine 
the auxiliary fields $M^i=M^i(\f,\bar{\f})$,
$$ \fracm{1}{2}M^i Z_{,i\bar{p}}+\bar{W}_{,\bar{p}}=0~. \eqno(3.5)$$
The bosonic scalar potential in components is thus given by 
$$V_{\rm scalar}(\f,\bar{\f}) = 
\left. \fracm{1}{2}M^i\Tilde{W}_{,i}\right|_{M=M(\f,\bar{\f})}~~,\eqno(3.6)$$
in agreement with ref.~\cite{spain}. In the next sect.~4 we study some 
examples, by restricting our general results to the case of a single chiral 
superfield, and then to the case of the $CP^1$ K\"ahler potential.

\section{Examples}

It is quite natural to begin with an example of a {\it single} chiral 
superfield, while keeping arbitrary both K\"ahler and scalar functions. In 
this case all the $\x$-integrations can be easily performed, e.g.,  by using 
eq.~(2.8) and the related identity obtained by differentiating eq.~(2.8) 
with respect to the parameter $a$,
$$ f(x+a) + f(x-a) =\int^{+1}_{-1}d\x\,f(x+\x a) + a\fracmm{\pa}{\pa x}
\int^{+1}_{-1}d\x\,\x f(x+\x a)~.\eqno(4.1)$$
Of course, one would get the same results by using the chiral reduction of
sect.~2 and the crucial identity (2.6) from the very beginning. We found some
cancellations amongst the bosonic terms, with the result
$$\eqalign{
L_{\rm bosonic} = & 
+\fracm{1}{2}\pa^{\m}\bar{\f}\pa_{\m}\bar{\f}\left[
K_{,\bar{\f}\bar{\f}}(\f+cM,\bar{\f})+K_{,\bar{\f}\bar{\f}}(\f-cM,\bar{\f})
\right] \cr
& + \fracm{1}{2}\bo\bar{\f}\left[
K_{,\bar{\f}}(\f+cM,\bar{\f})+K_{,\bar{\f}}(\f-cM,\bar{\f})\right] \cr
& + \fracmm{\bar{M}}{2c}\left[ K_{,\bar{\f}}(\f+cM,\bar{\f}) 
 -K_{,\bar{\f}}(\f-cM,\bar{\f}) \right] \cr
& + \fracmm{1}{2c}\left[ W(\f+cM) -W(\f-cM)\right]+\bar{M}
\fracmm{\pa\bar{W}}{\pa\bar{\f}}~~.\cr} \eqno(4.2)$$
The bosonic terms are to be supplemented by the following fermionic terms:
$$\eqalign{
L_{\rm fermionic} = & -\fracmm{1}{4c}\bar{\c}^2\left[ 
K_{,\bar{\f}\bar{\f}}(\f+cM,\bar{\f})-K_{,\bar{\f}\bar{\f}}(\f-cM,\bar{\f})
\right] \cr
& -\fracmm{i}{2cM}(\c\s^{\m}\bar{\c})\pa_{\m}\bar{\f}\left[
K_{,\bar{\f}\bar{\f}}(\f+cM,\bar{\f})-K_{,\bar{\f}\bar{\f}}(\f-cM,\bar{\f})
\right] \cr
& -\fracmm{i}{2cM}(\c\s^{\m}\pa_{\m}\bar{\c})\left[
K_{,\bar{\f}}(\f+cM,\bar{\f})-K_{,\bar{\f}}(\f-cM,\bar{\f})
\right] \cr
& -\fracmm{\bar{M}}{4cM}\c^2 
\left[ K_{,\f\bar{\f}}(\f+cM,\bar{\f})-K_{,\f\bar{\f}}(\f-cM,\bar{\f})
\right]  \cr
& -\fracmm{1}{4M}\c^2 \pa^{\m}\bar{\f}\pa_{\m}\bar{\f} 
\left[ K_{,\f\bar{\f}\bar{\f}}(\f+cM,\bar{\f})+K_{,\f\bar{\f}\bar{\f}}
(\f-cM,\bar{\f})\right]  \cr
& +\fracmm{1}{4cM^2}\c^2 \pa^{\m}\bar{\f}\pa_{\m}\bar{\f} 
\left[ K_{,\bar{\f}\bar{\f}}(\f+cM,\bar{\f})-K_{,\bar{\f}\bar{\f}}
(\f-cM,\bar{\f})\right]  \cr
& -\fracmm{1}{4M}\c^2\bo\bar{\f}
\left[ K_{,\f\bar{\f}}(\f+cM,\bar{\f})+K_{,\f\bar{\f}}
(\f-cM,\bar{\f})\right]  \cr 
& +\fracmm{1}{4cM^2}\c^2\bo\bar{\f}
\left[ K_{,\bar{\f}}(\f+cM,\bar{\f})-K_{,\bar{\f}}
(\f-cM,\bar{\f})\right]  \cr 
& +\fracmm{1}{8cM}\c^2\bar{\c}^2
\left[ K_{,\f\bar{\f}\bar{\f}}(\f+cM,\bar{\f})-K_{,\f\bar{\f}\bar{\f}}
(\f-cM,\bar{\f})\right]  \cr\
& -\fracmm{1}{4cM}\c^2\left[ W_{,\f}(\f+cM) - W_{,\f}(\f-cM)\right]
-\fracm{1}{2}\bar{\c}^2 \bar{W}_{,\bar{\f}\bar{\f}}~.\cr}\eqno(4.3)$$

In the case of the $CP^1$ K\"ahler potential 
$$ K(\f,\bar{\f}) =\ln ( 1+ \f\bar{\f})~~,\eqno(4.4)$$
the auxiliary field equation (3.5) gives rise to a quadratic equation on $M$,
whose roots are given by
$$ M= \fracmm{1\pm \sqrt{1+(2c\bar{\f}(1+\f\bar{\f})\bar{W}')^2}}{2c^2
\bar{\f}^2\bar{W}'}~~,\eqno(4.5)$$
where we have used the notation $\bar{W}'=\pa\bar{W}/\pa\bar{\f}\;$. Taking the
anticommutative limit $c\to 0$ implies that we should choose minus in 
eq.~(4.5). The scalar potential is given by eq.~(3.6), after substituting
eqs.~(3.4) and (4.5) overthere.

More examples and applications will be considered elsewhere \cite{nac3}.

\section{Conclusion}

A comparison to the NAC-deformed N=2 NLSM in {\it two} dimensions 
\cite{chandra,spain} is possible after dimensional reduction of our results
in sects.~3 and 4 by assuming $\pa_3=\pa_4=0$. As was already demonstrated in
ref.~\cite{spain}, the infinite series found in ref.~\cite{chandra} can be 
resummed in terms of the `minimally' deformed K\"ahler potential and 
superpotential in the sense of eqs.~(3.2a) and (3.4), plus some additional 
`non-minimal' terms with deformed coupling as in eq.~(3.2b), in {\it precise} 
agreement with our basic formulae (2.6) and (2.9).~\footnote{The split (2.6) 
of a NAC-deformed superpotential in the case of a single chiral superfield was
 found in ref.~\cite{nac1}. The smearing (2.9) and (2.10) of NAC-deformed 
K\"ahler potential and superpotential in the case of several chiral 
superfields was found in ref.~\cite{spain}.} As regards the infinite series 
found in refs.~\cite{buch,nbi} in four dimensions, those results seem to be 
very complicated to allow us a direct comparison. An explicit resummation is 
necessary not only (i) for comparison but also (ii) checking the locality 
of the deformed action in spacetime, (iii) verifying the auxiliary fields to 
be still non-propagating, and (iv) ultimately solving the auxiliary field 
equations. We believe that our results in this Letter are useful for all those
 purposes, because we addressed the most general case, formulated our results 
in a compact and transparent form, and offered a clear `short cut' for their 
easy derivation. 

As is well known in the theory of NLSM (see e.g., ref.~\cite{nlsm}), the 
so-called quotient construction (or gauging isometries of the NLSM target 
space) can be used to represent some NLSM with homogeneous target spaces as the
gauge theories. It was used in ref.~\cite{chuo} to construct the NAC-deformed 
supersymmetric NLSM in four dimensions with the $CP^n$ target space, by 
combining the quotient construction with the results of ref.~\cite{sei} about 
the NAC-deformed supersymmetric gauge theories. As is clear from our results
about generic NAC-deformed NLSM in sect.~3, the NAC deformation of a K\"ahler 
potential is controlled by the auxiliary fields entering the deformed K\"ahler
potential in the highly non-linear way. The auxiliary fields are determined by
their algebraic equations of motion that are also dependent upon a 
superpotential, even when all fermions are ignored. As a result, the NAC
deformation of K\"ahler geometry is controlled by a scalar superpotential! It
is in drastic contrast with the standard supersymmetric (K\"ahler) NLSM in 
undeformed superspace, whose target space geometry is unaffected by a scalar 
superpotential. In the absence of a scalar superpotential, $W=\bar{W}=0$,  we
found that the NAC deformation $c\neq 0$ does {\it not\/} affect the 
supersymmetric $CP^1$ NLSM action at all, in agreement with ref.~\cite{chuo}. 

A detailed analysis of the actions (3.1), (4.2) and (4.3), including the 
deformed supersymmetric $CP^n$ NLSM, will be given elsewhere \cite{nac3}.

\newpage

\section*{Acknowledgements}

The authors are grateful to Y. Kobayashi for discussions, B. Chandrasekhar, 
A. Kumar and A. Lerda for correspondence, and the referee for his careful 
reading of our manuscript and useful suggestions.

\end{document}